\documentstyle[11pt,epsfig]{article}
\begin{document}
\setlength{\baselineskip}{.222in}
\newcommand{\nc}{\newcommand}
\newcommand{\bi}{\bibitem}
\newcommand{\beq}{\begin{equation}}
\newcommand{\eeq}{\end{equation}}
\newcommand{\be}{\begin{eqnarray}}
\newcommand{\ee}{\end{eqnarray}}
\def\lapprox{\mathrel{\mathop
  {\hbox{\lower0.5ex\hbox{$\sim$}\kern-0.8em\lower-0.7ex\hbox{$<$}}}}}
\def\gapprox{\mathrel{\mathop
  {\hbox{\lower0.5ex\hbox{$\sim$}\kern-0.8em\lower-0.7ex\hbox{$>$}}}}}

\begin{center}
\vglue .06in
{\Large \bf {Decaying neutrino and a high cosmological baryon density}}\\
\bigskip
S.H. Hansen\footnote{e-mail: {\tt sthansen@fe.infn.it}} and
F.L. Villante\footnote{e-mail: {\tt villante@fe.infn.it}}
 \\[.05in]
{\it{Dipartimento di Fisica and Sezione INFN di Ferrara\\
Via del Paradiso 12,
44100 Ferrara, Italy}
}
\\[.40in]
\end{center}

\begin{abstract}
The low second acoustic peak in the recent Boomerang data may indicate
a cosmological baryon density which is larger than allowed by standard
big bang nucleosynthesis. 
We show that the decay of the tau-neutrino: $\nu_\tau \rightarrow
\nu_e + \phi$, where $\nu_e$ is the electron neutrino and $\phi$ is a 
scalar, essentially can assure
agreement between BBN calculations and light element observations
for a large baryon density.
\end{abstract}    


\section{Introduction}
The Boomerang experiment recently measured the
angular power spectrum of the cosmic microwave background up to
$l=600$~\cite{boom}. Accordingly to what is expected for a flat Universe,
the data show a peak in the power spectrum at $l=197\pm6$. At the same
time, however, the data seem to indicate that the second acoustic peak
is rather low~\cite{first}, which may be an indication of a high 
baryon number~\cite{tegmark} (see also~\cite{Hu:1995uz,white}) 
(say e.g. $\Omega_b h^2 \sim 0.03$, 
where $h$ is the Hubble constant in units $100 \, {\rm km \, s}^{-1} \, 
{\rm Mpc}^{-1}$). 
This simple conclusion, however, immediately leads to disagreement with the
well established Big Bang Nucleosynthesis (BBN), which predicts 
$ \Omega_b h^2 \approx 0.019 \pm 0.0024$~\cite{tytler}.
Although this discrepancy is still very preliminary, it is interesting
to investigate specific models which can reconcile 
BBN with a high baryon density.

The problem is the following. In the standard BBN scenario there is only one
free parameter, namely the baryon to photon number ratio,
$\eta = n_B/n_\gamma$, which is related to the baryon density according 
to $\eta\simeq 2.68 \cdot 10^{-8} \Omega_b h^2 $. Observationally 
$\eta_{10} = \eta \cdot 10^{10}$ has 
long been known to be in the interval $1$ {\small $\lapprox$}
$\eta _{10}$ {\small $\lapprox$} $10$, 
and the recent
deuterium measurements favor $\eta_{10} \approx 5$.
When one increases $\eta_{10}$ the helium abundance increases slightly while 
the deuterium abundances decreases rapidly:
\be
\eta_{10} \nearrow \, \, \, \, {\rm implies} \, \, \, \, 
Y_{he} \nearrow \, \, \, \, {\rm and} \, \, \, \, D/H \searrow
\ee
It is then clear that a high baryon number leads to a too
low deuterium prediction and a too high helium-4 value. 
Now, with a lower deuterium abundance one must seek a method of increasing the
deuterium, to achieve agreement with observations. 
This is easily found; one could simply increase the effective
number of massless degrees of freedom. 
For two different value of $\eta$ one can achieve the 
same deuterium abundance by varying the energy density
(expressed through $\Delta N_{eff} = N_{eff} - 3$). According to \cite{franc}, 
the needed $\Delta N_{eff}$ is roughly:
\be
\Delta N_{eff} \approx \frac{1}{0.03} \cdot
 {\rm log_{10}} \left( \frac{\eta_2}{\eta_1}  \right)
\label{EF}
\ee
and we thus see, that if $\eta_{10}$ is 7 (9) instead of 5, then we need
additional energy density corresponding to 5 (8.5) extra neutrinos.
Increasing the energy density (which effectively corresponds to 
adding new particles), however, also affects the helium abundance. 
An increase in $N_{eff}$ correspond to a variation
$\Delta Y_{he} \approx 0.013\Delta N_{eff}$ in the helium abundance and the
observational data leave room for not more than one extra neutrino.
We thus conclude that even if one can make deuterium 
calculations agree with observations by increasing $N_{eff}$, 
then the helium predictions will be in strong
disagreement with observations. The goal of this Letter is to 
point out a specific model, which may solve this apparent problem.

\section{A possible solution}
As mentioned above, a low second peak in the CMB power spectrum 
may be explained by a high baryon number. 
\begin{figure}[htb]
\begin{center}
  \leavevmode
  \hbox{
    \epsfysize=2.7in
    \epsffile{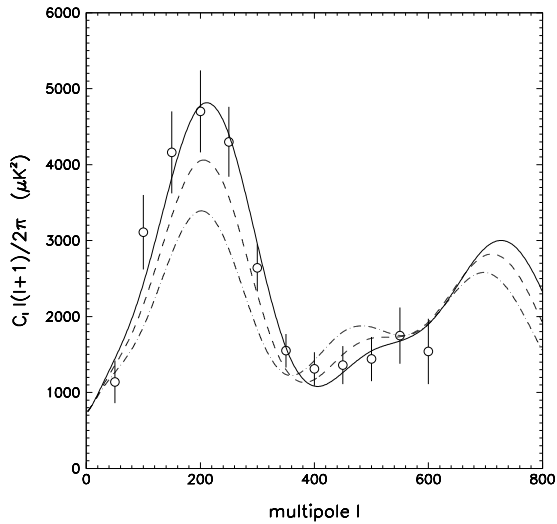}}
\end{center}
\caption{A comparison of the Boomerang data with theoretical predictions
for 3 flat CDM models with varying $\Omega_b$.
We have chosen values $h=0.68, \Omega_{tot} = 1, \Omega_\Lambda =0,
Y_{he} = 0.24$.
The various lines correspond to 
$\eta_{10} = 5$ (dash dotted),  $\eta_{10} = 7$ (dashed)
and $\eta_{10} = 9$ (solid).  There is no scalar tilt, no tensors and we 
neglect reionization.} 
\label{figpowersp}
\end{figure}
In fig.~1 we present a CMB power spectrum together with the Boomerang 
data~\footnote{The power spectra were made with the 
{\tt CMBFAST} code~\cite{cmbfast}.}.
With this figure we only want to illustrate that when 
varying  $\Omega_b$ in a flat, CDM Universe 
($\Omega_\Lambda = 0$) one gets a better agreement with a 
low second peak when $\Omega_b$ is high
(this is well known, see e.g. refs.~\cite{julien,Hu:1995uz}). 
In particular we find (for the chosen parameters) that $\eta_{10}=9$
gives a power spectrum in visually rather fair agreement with
the Boomerang data.
Such a high baryon number could be in agreement with deuterium
observations, if one at the same time allows for a large $N_{eff}$,
however, both effects lead to a larger helium abundance, 
in disagreement with observations.
Since we have used all the free parameters of BBN, we can only lower
the helium predictions by introducing new physics.

The helium abundance is determined by the electron neutrinos, $\nu_e$,
which govern the freeze-out of the $n-p$ reaction.
In the standard BBN calculations it is assumed that the electron neutrinos
have a Fermi Dirac thermal distribution. Any changes
to the distribution function of the neutrino will alter the predicted 
abundances. As is well known (see e.g.~\cite{dhs}), 
if one adds neutrinos at the 
high energy tail of the distribution function, then the final helium abundance
will be higher, whereas more low-energy $\nu_e$ will lower $Y_{he}$.
Likewise more thermal $\nu_e$ will lead to a decrease in helium.

More $\nu_e$ could be achieved in several ways. If one has mixing between a
massive tau-neutrino, $\nu_\tau$, and a muonic-neutrino, $\nu_\mu$, then the 
3-body decay: $\nu_\tau \rightarrow \nu_\mu+\nu_e+\bar{\nu_e}$, would be 
possible. The lifetime for this decay would go like:
\[
\tau_{\nu_\tau \rightarrow 3\nu} \approx \tau_\mu \left(  
\frac{m_\mu}{m_{\nu_\tau}}
\right)^5/{\rm sin}^22\theta
\]
where $\tau_\mu = 2.2 \cdot 10^{-6}$ sec and $m_\mu=106$ MeV is 
the muon decay time and mass, and sin$2\theta$ is
the mixing angle between the two neutrinos. For a big mixing angle and a
mass of the tau-neutrino of the order 1 MeV, one gets a lifetime of the order
$10^4$ sec, which would have very little influence on BBN. 
Recently was considered the possibility of adding a chemical potential both 
for the $\nu_e$ and for the $\nu_\tau$ in such a way, that BBN can have 
successful predictions even with large $\Omega_b$~\cite{julien,kang} (see 
also~\cite{kinney}).
This is achieved by letting the degeneracy parameter $\xi_{\nu_\tau} \sim 1$ 
provide more energy density, and a positive $\xi_{\nu_e} \ll 1$ to lower the 
$n-p$ ratio.
Also sources of Ly $\alpha$ resonance radiation (e.g. from hot stars or 
quasars), if present around $z \sim 1000$,
could delay recombination and hence lead to a lower second peak~\cite{peebles}.

Another possibility, which we will consider below, 
is the decay of a massive tau-neutrino into an electron
neutrino and a scalar:
\be
\nu_\tau  \rightarrow \nu _e + \phi
\ee
where $\phi$ is light (or massless). This scalar boson could
possibly be a Majoron~\cite{majorons}, and the effect on nucleosynthesis
of this decay was calculated accurately in ref.~\cite{dhps}~\footnote{In
refs.~\cite{turner1,turner2} was considered the effects of both this decay 
$\nu_\tau  \rightarrow \nu _e + \phi$ and also 
$\nu_\tau \rightarrow 3 \nu_e$ on all light elements for a heavy and 
longliving $\nu_\tau$.}.
The effects of this decay on BBN are the following. 
First, because the tau-neutrino is massive, it will contribute more 
to the energy density of the Universe, leading to a 
slightly increased  deuterium abundance.
For long lifetimes the mass effect of the $\nu_\tau$ may boost the energy
density of the Universe corresponding to up to 
$\Delta N_{eff} = 7$ extra neutrinos (see fig.~1a
in \cite{dhps}).

The variation in the helium abundance does not follow directly
from the increase in the energy density. This is because for helium
the energy distribution of the decay products (the $\nu_e$) 
play a relevant role.
In particular, for a wide range of values for $m$ and $\tau$,
the net effect could be a decrease in the helium abundance
as seen in fig.~2, and described in ref.~\cite{dhps}\footnote{The data
used in fig.~2 are taken from the web site 
{\tt http://tac.dk/\~{}sthansen/decay/}.}.
\begin{figure}[htb]
\begin{center}
  \leavevmode
  \hbox{
    \epsfysize=2.7in
    \epsfxsize=2.7in
    \epsffile{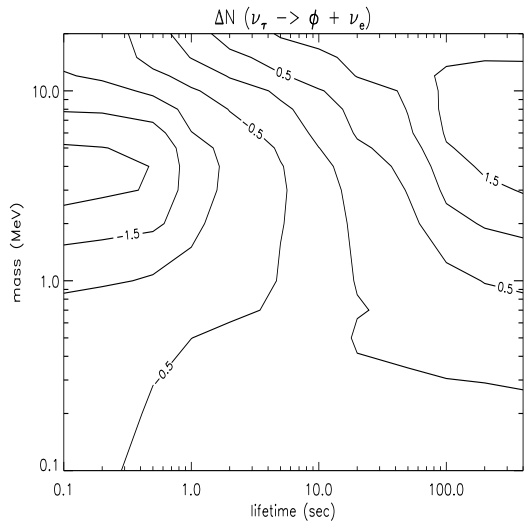}}
\end{center}
\caption{$\Delta N_\nu$ contour lines found from helium-4, as a function of
$\nu_\tau$ mass and lifetime for $\eta _{10} = 9$. The contours showed 
correspond to $\Delta N_\nu =-2,-1.5,-1,-0.5,0,0.5,1,1.5$. One notes, that for
small lifetime or small mass 
one can achieve a decrease in the helium abundance.} 
\label{contour}
\end{figure}
Specifically we find, that by choosing a mass, 
$m_{\nu_\tau}$, of a few MeV, and
a lifetime, $\tau$, of a few seconds, 
one  has both more relativistic energy density
than provided with 3 normal neutrinos, and at the same time the helium 
abundance is lowered substantially. For e.g. $m=4$ MeV and $\tau =2$ sec,
we see that $Y_{he}$ is lowered by $\sim 10\%$, which {\it for what 
concerns helium production} correspond to 
$\Delta N_{\nu} \approx -1$~\footnote{Let us for the reference  clarify
that $\Delta N_{eff}$ parameterizes the change in energy density,
whereas  $\Delta N_\nu$ describes the change in primordial helium-4.}.
A region in the $(m_{\nu_\tau}, \tau_{\nu_\tau})$-parameter space
even provides one with such a low helium abundance that there is freedom
to add two sterile neutrinos (see small lifetimes on fig.~2),
which can increase the deuterium 
abundance further, if needed.

The two regions in $(m_{\nu_\tau}, \tau_{\nu_\tau})$-parameter space
described above, namely the long lifetime region ($\tau \gapprox$ few sec)
in which deuterium is 
increased ($\Delta N_{eff} > 0$),
 and the $\Delta N_{\nu}<0$ region in which helium
is decreased (see fig.~2) only partially overlap.
This means that the region where one can both decrease helium and at 
the same time increase
deuterium
is fairly small. With an approximate analysis, where we optimize 
$\Delta N_{eff}-\Delta N_{\nu}$ in the overlap region and use 
eq.~\ref{EF}, it seems 
only possible to allow for $\eta_{10}$ smaller than 7.
It is interesting to note, that in this overlap region the decaying neutrino
also leads to a decrease in the amount of lithium, which will be needed to
conform with lithium observations~\footnote{We are grateful to P.~Bonifacio
for comments on this point.}.

Let us mention that late decaying $\nu_\tau$ with $m>3.6$ MeV
may also produce high energy $\nu_e$, which further can increase the deuterium 
abundance through the reaction: 
$\nu_e + p \rightarrow n +e^+$~\cite{scherrer,dhps}.
This effect become more efficient for high $\eta$.

Finally, 
it is worth mentioning that the Boomerang data together with BBN deuterium 
arguments can give a fairly interesting upper limit on $\eta$. It was
found in ref.~\cite{hannestad} from the Boomerang data that even when allowing 
$\Omega_b$ to vary within a large region one gets a $2\sigma$ bound
on the relativistic energy density, $N_{eff} < 13$. Now, using 
eq.~(\ref{EF}) one can translate this into a bound on $\eta$, namely 
$\eta_{10} < 12.5$. This translation is not quite safe for two reasons.
First, in~\cite{hannestad} $\Omega_b h^2$ was only allowed to vary up to
$0.03$ ($\eta_{10} = 9$) which is smaller than 
the bound just found and hence the extrapolation to $\eta_{10} < 12.5$ is 
strictly speaking not justified~\footnote{We are grateful to S.~Hannestad
for comments on this point.}, and second,
the formula in eq.~(\ref{EF}) is  approximate. It could be interesting 
to do a careful analysis.

\section{Conclusion}
We have shown, how a specific model with a 
decaying massive tau-neutrino can make BBN calculations
for the light element abundances agree with the observations in a Universe 
with as high baryon number as $\eta_{10}=7$.
Such a high baryon number may be needed to explain
a lower second peak in the CMB power spectrum as seen in the recent 
Boomerang data.

The scenario with a high $\Omega_b$ naturally  predicts a high $3^{\rm rd}$
peak, and can hence easily be excluded by future CMB observations. On the
other hand, should the future CMB experiments find a high $3^{\rm rd}$ peak,
then one must distinguish between the various models. One can distinguish a 
high $\Omega_b$ scenario from the delayed recombination picture suggested in
ref.~\cite{peebles}, since high $\Omega_b$ will lower the diffusion damping, 
and hence the  $3^{\rm rd}$ peak should be higher in this case than in 
the models proposed in~\cite{peebles}.
Distinguishing between a decaying neutrino and a chemical 
potential~\cite{julien} is more difficult, and one would probably need 
refined observations of other light elements like helium-3 and lithium.

\section*{Acknowledgment}
It is a pleasure to thank A.D. Dolgov for kind suggestions and comments.

\end{document}